\documentclass[letter]{IEEEtran}
\usepackage{graphicx}
\usepackage{fancyhdr}
\usepackage{makeidx}
\usepackage{amssymb,amsmath}
\usepackage{enumerate}
\newcommand{\bd}{\begin{document}}
\newcommand{\ed}{\end{document}}
\newcommand{\bc}{\begin{center}}
\newcommand{\ec}{\end{center}}
\newcommand{\vs}{\vspace}
\newcommand{\hs}{\hspace}
\newcommand{\beq}{\begin{equation}}
\newcommand{\eeq}{\end{equation}}
\newcommand{\beqs}{\begin{eqn*}}
\newcommand{\eeqs}{\end{eqn*}}
\newcommand{\bq}{\begin{quote}}
\newcommand{\eq}{\end{quote}}
\newcommand{\lb}{\linebreak}
\newcommand{\mb}{\makebox}
\newcommand{\fb}{\framebox}
\newcommand{\mc}{\multicolumn}
\newcommand{\ben}{\begin{enumerate}}
\newcommand{\een}{\end{enumerate}}
\newcommand{\bit}{\begin{itemize}}
\newcommand{\eit}{\end{itemize}}
\newcommand{\ov}{\overline}
\newcommand{\un}{\underline}
\newcommand{\lt}{\left}
\newcommand{\rt}{\right}
\newcommand{\ba}{\begin{array}}
\newcommand{\ea}{\end{array}}
\newcommand{\beqa}{\begin{eqnarray}}
\newcommand{\eeqa}{\end{eqnarray}}
\newcommand{\beqas}{\begin{eqnarray*}}
\newcommand{\eeqas}{\end{eqnarray*}}
\newcommand{\bfg}{\begin{figure}}
\newcommand{\efg}{\end{figure}}
\newcommand{\pad}{\partial}
\newcommand{\nn}{\nonumber}
\newcommand{\la}{\leftarrow}
\newcommand{\ra}{\rightarrow}
\newcommand{\lgla}{\longleftarrow}
\newcommand{\lgra}{\longrightarrow}
\newcommand{\La}{\Leftarrow}
\newcommand{\Ra}{\Rightarrow}
\newcommand{\Lra}{\Leftrightarrow}
\newcommand{\Lgla}{\Longleftarrow}
\newcommand{\Lgra}{\Longrightarrow}
\renewcommand{\a}{\alpha}
\renewcommand{\b}{\beta}
\newcommand{\g}{\gamma}
\newcommand{\G}{\Gamma}
\renewcommand{\d}{\delta}
\newcommand{\D}{\Delta}
\newcommand{\e}{\epsilon}
\newcommand{\eps}{\epsilon}
\newcommand{\s}{\sigma}
\renewcommand{\l}{\lamda}
\newcommand{\m}{\mu}
\newcommand{\n}{\nu}
\renewcommand{\S}{\Sigma}
\newcommand{\p}{\pi}
\newcommand{\om}{\omega}
\newcommand{\Om}{\Omega}
\newcommand{\tri}{\triangle}
\newcommand{\ti}{\times}
\newcommand{\f}{\frac}
\newcommand{\ds}{\displaystyle}
\newcommand{\bm}[1]{\mb{{\boldmath $#1$}}}
\newcommand{\alter}[2]{\lt\{ \ba{ll}#1 \\ #2 \ea \rt.}
\newcommand{\alt}[4]{\lt\{ \ba{ll}#1 & \mb{if \, \,}#2 \\ #3 & \mb{}#4 \ea
    \rt.}
\newcommand{\altn}[4]{\lt\{ \ba{rl}#1 & \mb{if \, \,}#2 \\ #3 & \mb{}#4 \ea
    \rt.}
\newcommand{\altif}[4]{\lt\{ \ba{ll}#1 & \mb{if \, \,}#2 \\ #3 &
\mb{if \, \,}#4 \ea \rt.}
\newcommand{\altnif}[4]{\lt\{ \ba{rl}#1 & \mb{if \, \,}#2 \\ #3 &
\mb{if \, \,}#4 \ea \rt.}
\newcounter{algc}
\newcounter{romc}
\newcounter{Alphc}
\newcommand{\bl}{\begin{list}{{\it Step} ~\arabic{algc}~:} {\usecounter{algc}
                \setlength{\topsep}{0pt} \setlength{\itemsep}{0pt}}}
\newcommand{\el}{\end{list}}
\newcommand{\blr}{\begin{list}{~\roman{romc}~:} {\usecounter{romc}
                \setlength{\topsep}{0pt} \setlength{\itemsep}{0pt}}}
\newcommand{\elr}{\end{list}}
\newcommand{\bla}{\begin{list}{~\Alph{Alphc}~:} {\usecounter{Alphc}
                \setlength{\topsep}{0pt} \setlength{\itemsep}{0pt}}}
\newcommand{\ela}{\end{list}}

\newtheorem{theorem}{Theorem}

\begin{document}
\bstctlcite{IEEEexample:BSTcontrol}
\title{Underlap Optimization in HFinFET in Presence of Interface Traps}
\author{Kausik Majumdar$^{1*}$, Rajaram S. Konjady$^2$, Raj Tejas S.$^3$ and Navakanta
Bhat$^1$\\
$^1$Department of Electrical Communication Engineering and Center
for Nanoscience and Engineering, Indian Institute of
Science, Bangalore 560012, India.\\
$^2$Oracle India Pvt. Ltd., IBC Knowledge Park, Bannerghatta road,
Bangalore 560029, India.\\
$^3$Electrical Engineering and Computer Science, University of
Michigan Ann Arbor, MI 48109, USA.\\
$^*$Corresponding author, Email: kausik@ece.iisc.ernet.in.}
\date{}
\maketitle {\abstract In this work, using 3D device simulation, we
perform an extensive gate to source/drain underlap optimization for
the recently proposed hybrid transistor, HFinFET, to show that the
underlap lengths can be suitably tuned to improve the on-off ratio
as well as the subthreshold characteristics in an ultra-short
channel n-type device without significant on performance
degradation. We also show that the underlap knob can be tuned to
mitigate the device quality degradation in presence of interface
traps. The obtained results are shown to be very promising when
compared against ITRS 2009 performance projections as well as
published state of the art planar and non-planar Silicon MOSFET data
of comparable gate lengths using standard benchmarking techniques.}

\section{Introduction}
The challenge to scale bulk MOSFET is ever increasing, particularly
beyond 22 nm technology node, due to significantly large short
channel effects and it gradually becomes essential to get equipped
with the alternate technologies to continue CMOS scaling
\cite{itrs09}. FinFET is one of the options which provides excellent
scalability due to its non-planar
structure\cite{ykc01}-\cite{fly06}. Recently, a hybrid architecture
of HEMT \cite{k07}-\cite{d05} and FinFET, called HFinFET, has been
proposed to obtain high performance as well as improved short
channel effects \cite{mmbj10}. The design relies on the improved on
performance from an HEMT-like on operation coupled with good short
channel control due to a FinFET-like multi-gate non-planar
structure.

The aim of this work is to investigate further scalability of
HFinFET using gate to S/D underlap length \cite{jgf03}-\cite{c07} as
a tuning knob and we show that without much performance degradation,
the on-off ratio as well as the subthreshold slope of an HFinFET can
be improved significantly for channel length down to $10$nm. We also
investigate the effect of channel-gate insulator interfacial traps
on device performance. This is very important because the device has
III-V channel material and the current technology does not offer an
excellent interface with gate dielectric \cite{gh08}-\cite{zh09}.
Finally, we show that the underlap parameter can be tuned to largely
mitigate the interface trap related issues, while meeting the
performance projection by ITRS 2009 \cite{itrs09}.

\section{Device Structure and Simulation Method}
The top view of the schematic diagram of a HFinFET and its cross
section along the dotted line $CC^\prime$ are shown in Fig.
\ref{fig:schematic}. The channel (Fin) is of
In$_{0.53}$Ga$_{0.47}$As sitting on an insulator. The hard mask on
the top is thick enough so that the effect of the top gate can be
neglected. A barrier layer, having a conduction band offset, is
sandwiched between the channel and the hard mask. During on
condition, the device electrostatics allows the delta doped layer in
the barrier layer to supply carriers to the channel. During off
state, these carriers can be pulled out of the channel by the
application of appropriate bias at the side gates, separated from
the channel by the gate dielectric. This combined effect of HEMT and
FinFET allows the device to obtain good on as well as off
performance. \bfg[htbp!]
\vs{-0.2in} \hs{-0.2in}
\includegraphics[width=3.8in,height=4in]{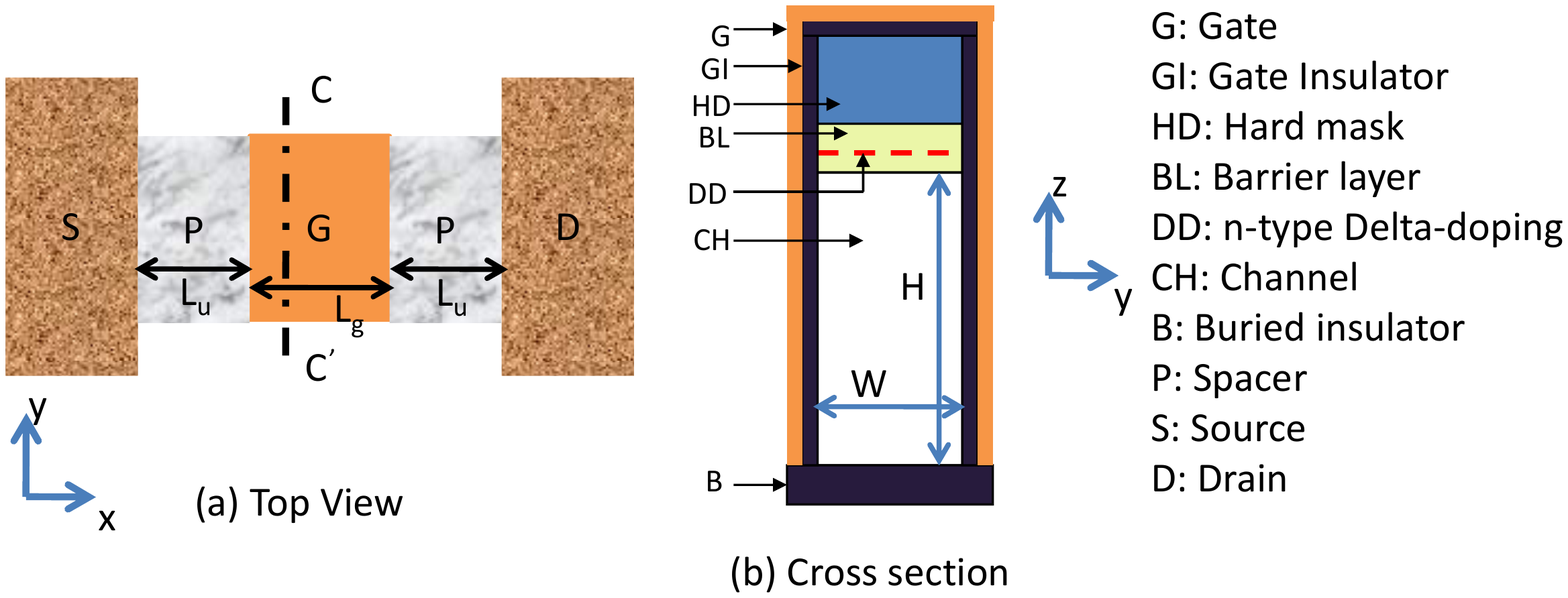}
\vs{-2in} \caption{The (a) top view and (b) cross-section along the
$CC^\prime$ of an HFinFET with the channel along the $x$ direction.
The different layers of the transistor are labeled in the
figure.}\label{fig:schematic}
\efg The metal gate work function ensures flatband on condition
\cite{p06}, which, coupled with the low carrier effective mass of
the undoped channel, pushes the transistor close to the ballistic
limit at shorter channel lengths. We have employed an effective mass
based 3-D Poisson-Schrodinger solver coupled with ballistic
transport \cite{kn94}-\cite{mbmj10} to simulate the device. More
details about the device operation and the simulation method are
explained in ref. \cite{mmbj10}. In this work, we assume the
equivalent oxide thickness (EOT) of the gate dielectric to be $1$nm.
Assuming use of high-k dielectric which allows thicker physical
oxide thickness \cite{gh08}-\cite{zh09}, the gate leakage has been
assumed to be negligible and has not been taken into consideration
in the simulation. The doping of the n-type delta doped layer is
assumed to be $2\times10^{12}$cm$^{-2}$, which supply electrons in
the channel. We consider the gate length ($L_g$) as $10$nm and
$15$nm with three different fin widths: $5$nm, $7.5$nm and $10$nm.
The fin height ($H$) is assumed to be $15$nm for all the cases. We
assume the supply voltage to be $0.7$V. To obtain a realistic
estimate on the device performance, we assume series source ($R_s$)
and drain resistance ($R_d$), each of which being
$200$$\Omega$$\mu$m. In the following, we use the four performance
criteria proposed in \cite{c05}, namely, (1) intrinsic gate delay
($\tau=CV/I$), (2) energy-delay product ($E.\tau=CV^2.CV/I$), (3)
subthreshold slope ($S$) and (4) intrinsic gate delay versus on-off
ratio, to benchmark the device performance. The performance
predictions are compared against ITRS 2009 projections \cite{itrs09}
as well as Si planar and non-planar devices of similar gate lengths
\cite{by02,c05,rc03}.

\section{Underlap Optimization}
In this work, we assume a symmetric gate-to-source and gate-to-drain
underlap ($L_u$). With an increase in $L_u$, the drain to source
coupling reduces significantly, particularly for a shorter channel,
which in turn gets reflected in the off state leakage as well as
subthreshold slope of the device. However, at the same time, the
underlap increases the series resistance of the channel, degrading
the on performance \cite{jgf03}-\cite{c07}. The results are shown in
Fig. \ref{fig:ulap_noDit} for two different channel lengths
($L_g$=$10$nm and $15$nm) with a fixed fin width of $W$=7.5nm. In
Fig. \ref{fig:ulap_noDit}(a), we observe a significant improvement
of on-off ratio with an increase in $L_u$, a clear indication of
improved off state control. As expected, the effect is more
prominent for the shorter channel device. However, as the underlap
length is increased beyond $\sim$$10$nm, the relative on-off
advantage saturates. For comparison, we have indicated the on-off
ratio obtained for a FinFET with $L_g$=$10$nm \cite{by02} and for a
planar MOSFET with $L_g$=$15$nm \cite{rc03}. \bfg[htbp!]
\vs{-0.1in} \hs{-0.25in}
\includegraphics[scale=0.35]{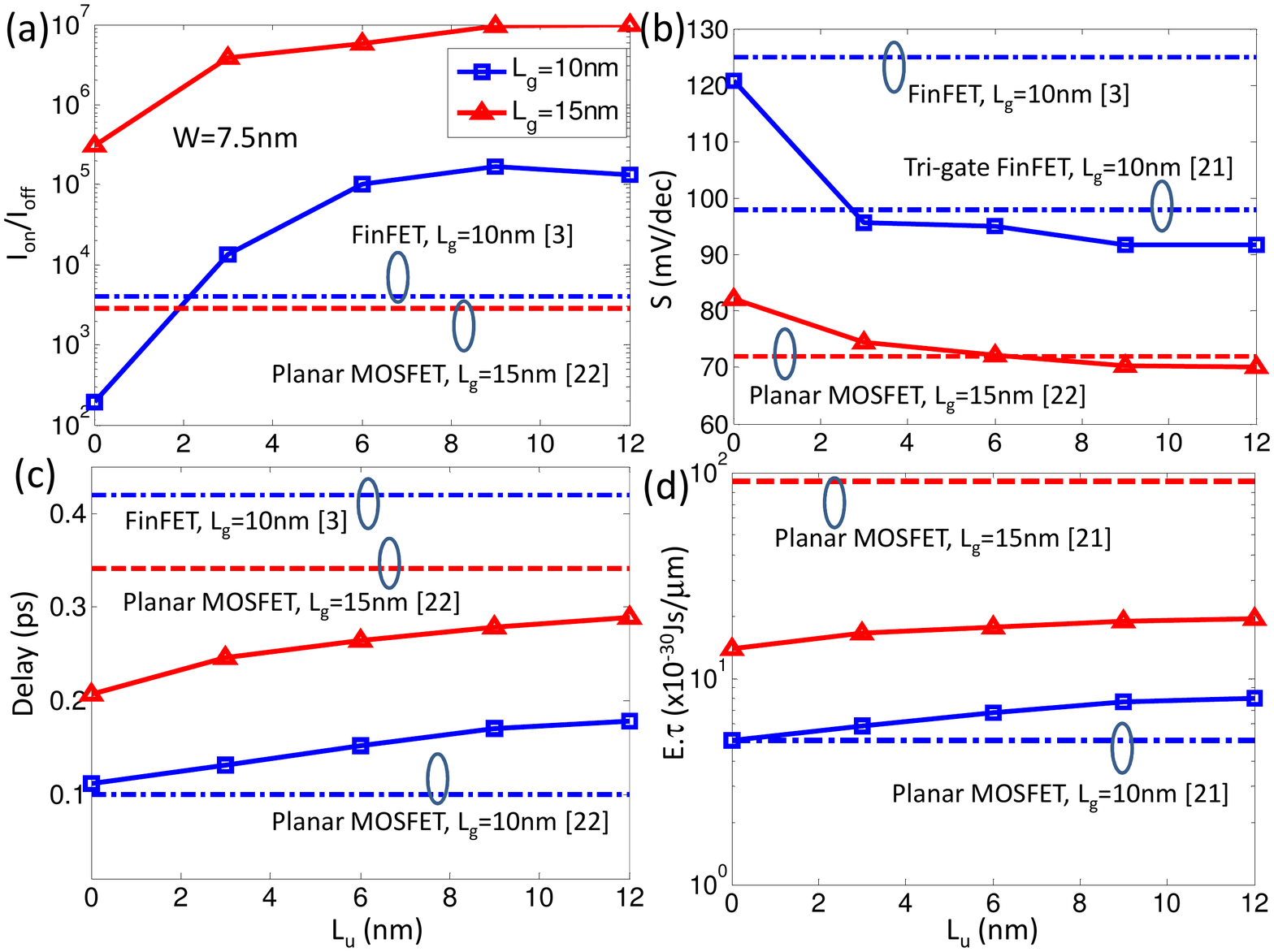}
\vs{-0.2in} \caption{(a) On-off ratio, (b) subthreshold slope, (c)
intrinsic gate delay and (d) energy-delay product of HFinFET as a
function of underlap for two different channel lengths, $10$nm
(squares) and $15$nm (triangles) with a width of $7.5$nm. The dotted
lines represent published planar and non-planar Si transistor data
for both $10$nm and $15$nm channel length. }\label{fig:ulap_noDit}
\efg
\bfg[htbp!]
\vs{-0.1in} \hs{-0.25in}
\includegraphics[scale=0.35]{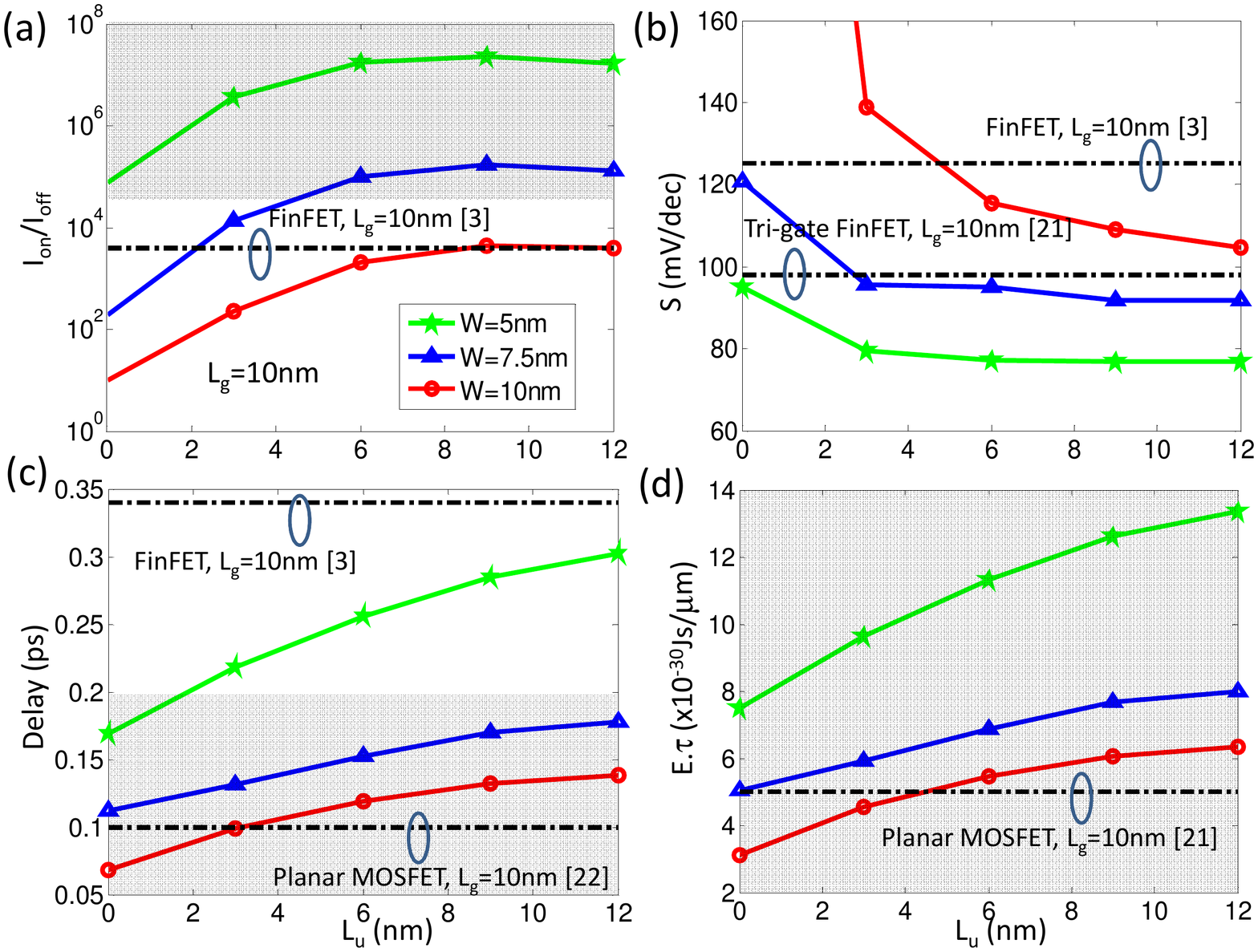}
\vs{-0.2in} \caption{(a) On-off ratio, (b) subthreshold slope, (c)
intrinsic gate delay and (d) energy-delay product of HFinFET as a
function of underlap with a channel length of $10$nm, for three
different fin widths, namely, $5$nm (stars), $7.5$nm (triangles) and
$10$nm (dots). The dotted lines represent published planar and
non-planar Si transistor data for $10$nm channel length. The shaded
areas represent those points that meet the performance projection of
$L_g \sim 10$nm by ITRS 2009.}\label{fig:ulap_noDit_w}
\efg We plot the variation of the subthreshold slope as a function
of $L_u$ in Fig. \ref{fig:ulap_noDit}(b) which shows a significant
improvement of subthreshold slope with an increase in $L_u$. In
particular, we are able to pull down the slope from $120$mV/dec to
$90$mv/dec for a $10$nm channel length device by altering the
underlap length. Here again, the sensitivity of the subthreshold
slope reduces significantly at relatively larger underlap. It is
also noticed that with a suitable underlap, HFinFET can provide
improved subthreshold slope as compared to published data on MOSFET
and FinFET of comparable gate lengths.

Fig. \ref{fig:ulap_noDit}(c) and (d) reflect the degradation of
intrinsic gate delay and the energy-delay product as a function of
the underlap length. It is observed that an increase in underlap
length from zero to $9$nm can cause a degradation of $50\%$ and
$35\%$ in the intrinsic delay, for $L_g$=$10$nm and $15$nm
respectively. Thus, it is important to carefully choose the underlap
length to meet the delay requirements. We notice a similar
degradation in the energy-delay product numbers as well with an
increase in underlap length.

In Fig. \ref{fig:ulap_noDit_w}, we plot the HFinFET characteristics
as a function of $L_u$ with different fin widths for a fixed $L_g$
of $10$nm. The shaded spaces indicate the regions which meet the
specifications prescribed by ITRS 2009. As expected, with an
increase in the fin width, the sensitivity of the characteristics
due to a change in the underlap increases. We note that, in Fig.
\ref{fig:ulap_noDit_w}(a), even with a large underlap, the
$W$=$10$nm case is not able to meet the ITRS projection for on-off
ratio. However, with smaller width, it can be met at suitable
underlap. The subthreshold slope in Fig. \ref{fig:ulap_noDit_w}(b)
has a strong degradation with increase in fin width due to stronger
source-drain coupling, which can be effectively reduced by
increasing the underlap. We clearly notice from Fig.
\ref{fig:ulap_noDit_w}(c) that even with sufficiently large
underlap, we are able to meet the ITRS projection for delay with
$W$$\geq$$7.5$nm. However, for smaller fin width ($W$=$5$nm), it is
difficult to meet the projection for any significant underlap.
However, as we see from \ref{fig:ulap_noDit_w}(d), the predicted
energy-delay products are safely within the limit of ITRS projection
for any width and underlap combination. This clearly indicates the
small switching charge in the device operation due to strong energy
quantization arising from both low carrier effective mass as well as
geometric confinement of the carrier wave function. All these
projected numbers compare very well against published planar and
non-planar Si MOSFET data.

\bfg[htbp!]
\vs{-0.2in} \hs{-0.2in}
\includegraphics[width=3.8in,height=3.6in]{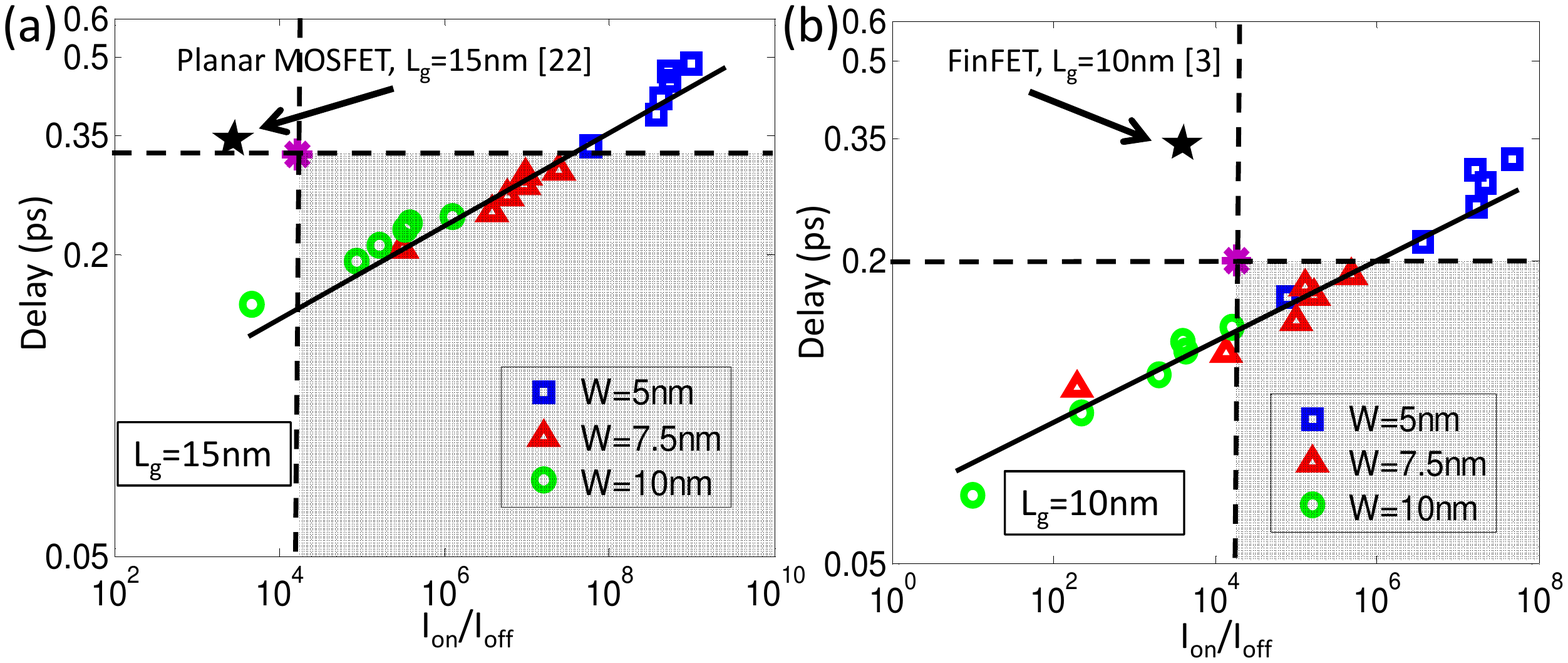}
\vs{-1.7in} \caption{Intrinsic gate delay versus on-off ratio for
(a) $L_g$=$15$nm and (b) $L_g$=$10$nm, for different underlap
lengths (0 to 12nm in steps of 3nm) and fin widths (5nm, 7.5nm and
10nm). For a given fin width, the increase in underlap tends to push
the points towards the top-right corner. The shaded areas represent
those points that meet the performance projection of comparable
channel lengths by ITRS 2009. We have also indicated a $15$nm planar
MOSFET \cite{rc03} and $10$nm FinFET \cite{by02} in the
plots.}\label{fig:onoff_delay_noDit}
\efg We show the transistor characteristics in the delay versus
on-off ratio space for both $L_g$=$15$nm and $10$nm in Fig.
\ref{fig:onoff_delay_noDit}(a) and (b) respectively. In both the
plots, for a given fin width, an increase in underlap shifts the
operating point towards the top right corner. Interestingly, for
$L_g$=$15$nm, there is no $W$=5nm point that falls inside the shaded
region that satisfies ITRS projection due to excessive intrinsic
delay and this remains true for any underlap. On the other hand, for
$L_g$=$10$nm, the operating points with $W$=$10$nm lie outside the
projected region due to poor on-off ratio. Also, note that, at
smaller fin width, there is comparatively larger spread of gate
delay and less spread in on-off ratio for different underlap
lengths. However, the converse is true for larger fin widths. As
expected, a less number of points satisfy ITRS projection for
$L_g$=$10$nm as compared to $L_g$=$15$nm.

Finally, in Fig. \ref{fig:off_delay_noDit}(a) and (b), the intrinsic
gate delay of the HFinFET is plotted as a function of off current,
for different fin widths and underlap lengths. It is shown that for
both the gate lengths, at a given fin width, an increased underlap
significantly improves the off current, though at the cost of
marginal delay degradation. \bfg[htbp!]
\vs{-0.2in} \hs{-0.2in}
\includegraphics[width=3.6in,height=3.3in]{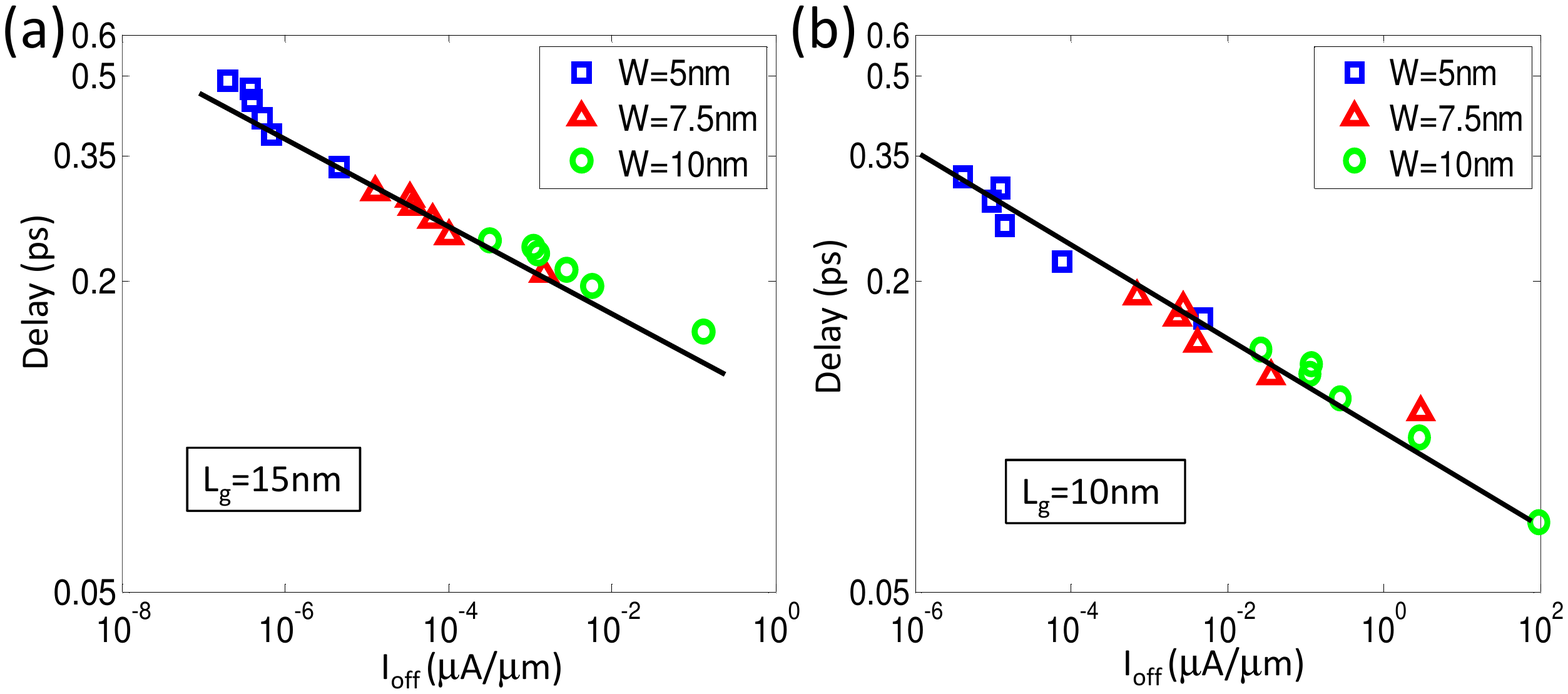}
\vs{-1.4in} \caption{Intrinsic gate delay versus off current for (a)
$L_g$=$15$nm and (b) $L_g$=$10$nm, for different underlap lengths (0
to 12nm in steps of 3nm) and fin widths (5nm, 7.5nm and 10nm). For a
given fin width, the increase in underlap increases the gate delay,
but at the same time reduces the off
current.}\label{fig:off_delay_noDit}
\efg

\section{Effects of Interface Traps}
In spite of intense research on improving the interface quality of
gate dielectric and III-V channel material \cite{gh08}-\cite{zh09},
to date, it remains one of the technological roadblocks that need to
be overcome. In this section, we assume a uniform distribution of
the interface trap density in the bandgap of the channel material at
the channel-dielectric interface and examine its effects on the
HFinFET characteristics. \bfg[htbp!]
\vs{-0.2in}
\hs{-0.2in}
\includegraphics[width=3.8in,height=3.6in]{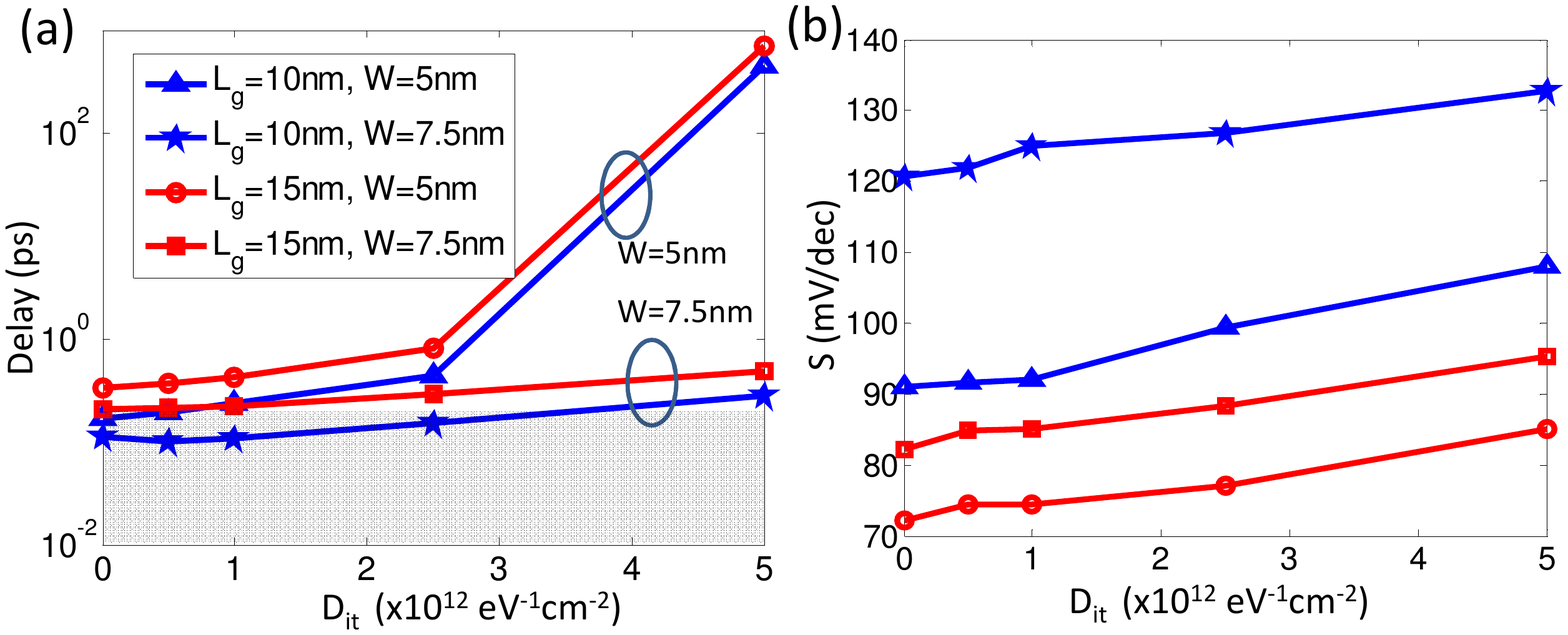}
\vs{-1.8in} \caption{(a) Intrinsic gate delay and (b) Subthreshold
slope of the HFinFET for different channel lengths and fin widths,
the underlap length being zero. The smaller fin widths tend to be
more prone towards increased trap density. The shaded area in (a)
represents the region meeting the performance projection of $L_g
\sim 10$nm by ITRS 2009.}\label{fig:dit_noUlap}
\efg

Importantly, the presence of interface traps increases the parasitic
capacitance which in turn reduces the fraction of useful (mobile)
channel charge in the total charge that is switched during device
operation \cite{mbmj10}. This effect is more prominent as we reduce
the fin width since this in turn reduces the quantum capacitance of
the channel due to stronger energy quantization arising from
geometrical confinement. This is explained in Fig.
\ref{fig:dit_noUlap}(a) where we plot the intrinsic gate delay as a
function of the trap density $D_{it}$ for different channel lengths
and fin widths with a zero underlap. It is observed that the gate
delay is less dependent on $D_{it}$ for relatively wider fins, but
for the narrower ones, the delay increases drastically. Larger
$D_{it}$ drives the operating point away from the region allowed by
the ITRS projection. In Fig. \ref{fig:dit_noUlap}(b), it is observed
that for a given channel length and fin width, the subthreshold
slope degrades almost linearly with an increase in $D_{it}$.

\bfg[htbp!]
\vs{-0.2in} \hs{-0.3in}
\includegraphics[width=4in,height=3.8in]{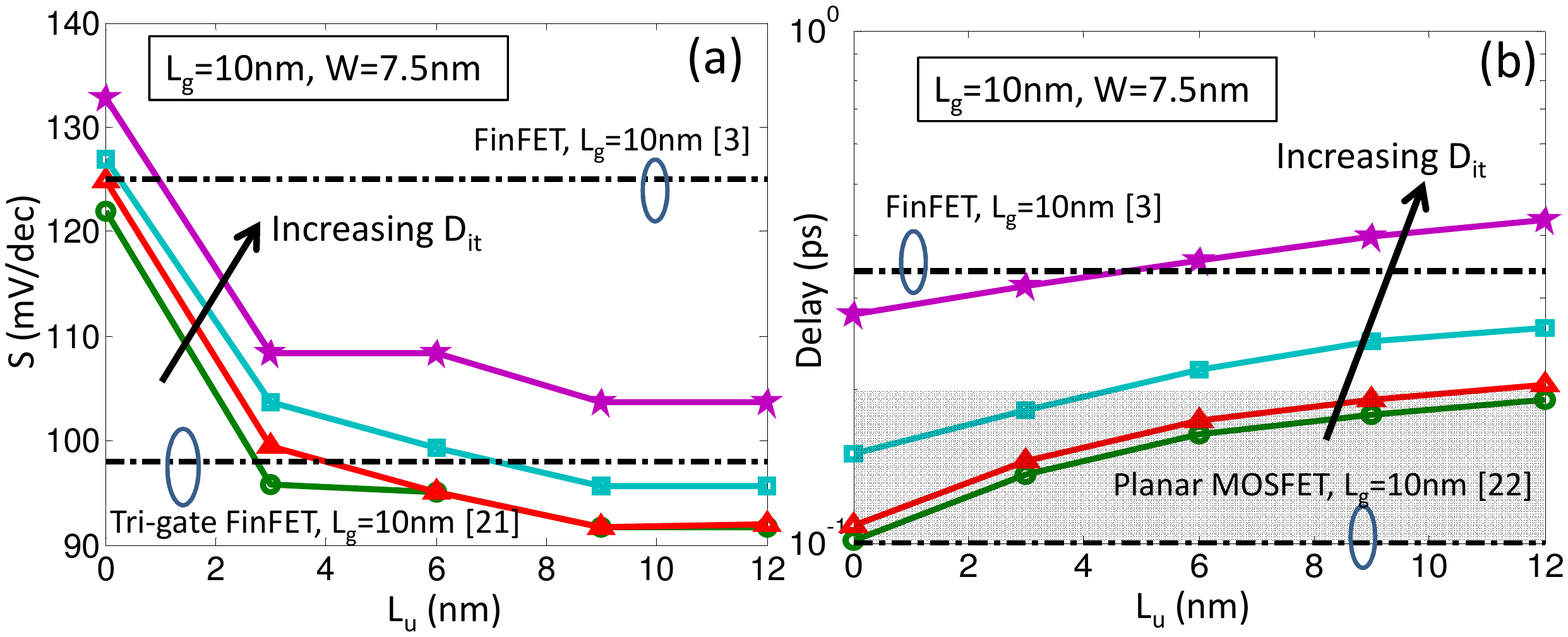}
\vs{-1.8in} \caption{The effect of increasing underlap length on (a)
Subthreshold slope and (b) intrinsic gate delay of HFinFET with
channel length of $10$nm and fin width of $7.5$nm, for trap
densities of $5\times10^{11}$, $10^{12}$, $2.5\times10^{12}$ and
$5\times10^{12}$eV$^{-1}$cm$^{-2}$. The subthreshold slope can be
significantly reduced by underlap, while meeting the ITRS delay
projection for $D_{it}$$\leq$$2.5\times10^{12}$eV$^{-1}$cm$^{-2}$.
}\label{fig:dit_ulap}
\efg We now show that this increase in subthreshold slope due to
increased $D_{it}$ can again be brought back to lower numbers by an
increase in underlap, as explained in Fig. \ref{fig:dit_ulap}(a). At
sufficiently larger underlap, the predicted subthreshold slope
values outperform the Silicon non-planar data \cite{by02,c05}, even
at significantly large trap density. The degradation in delay due to
the increased underlap is shown in Fig. \ref{fig:dit_ulap}(b) and
can be comfortably controlled to keep within ITRS limit, unless the
interface is extremely poor.

Finally, we have plotted the performance of HFinFET in presence of
$D_{it}$ and $L_u$ in the delay versus on-off ratio space in Fig.
\ref{fig:onoff_delay_dit}. We notice that there is no point that
meets the ITRS projections for
$D_{it}$=$5\times10^{12}$cm$^{-2}$eV$^{-1}$, even at large underlap
lengths. However, at smaller trap density, there is an allowed
window whose size strongly depends on the quality of the
channel-dielectric interface. Nonetheless, the underlap parameter
can be indirectly used to mitigate this technological challenge of
obtaining improved channel-dielectric interface.
\bfg[htbp!]
\vs{-0.2in}
\hs{-0.2in}
\includegraphics[scale=0.35]{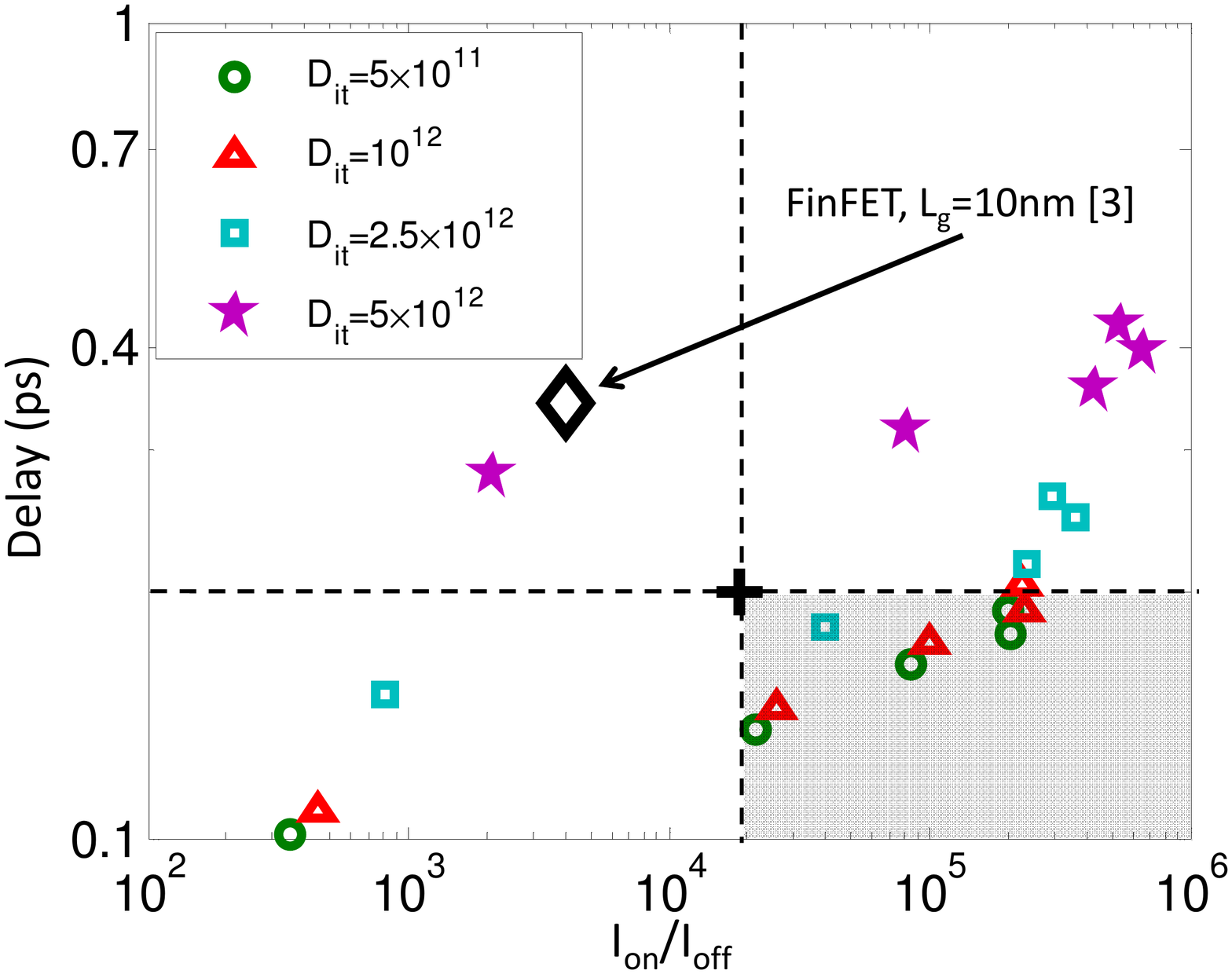}
\vs{-0.2in} \caption{Intrinsic gate delay versus on-off ratio
$L_g$=$10$nm and $W$=7.5nm, with different trap densities, the
underlap lengths varying from 0 to 12nm in steps of 3nm. With
increase in $D_{it}$, less number of points meet the ITRS 2009
projection. We have also indicated a performance of a published
$10$nm FinFET in the plot from ref.
\cite{by02}.}\label{fig:onoff_delay_dit}
\efg

\section{Conclusion}
To conclude, we have performed a 3-D simulation study to investigate
the effects of introducing a varying gate-source and gate-drain
underlap to show that it provides a unique way to meet either high
performance or low power requirements for the HFinFET. It has been
shown that the effect of the underlap strongly depends on the fin
width as well as the channel length of the transistor. The
performance degradation of HFinFET has been studied in presence of
traps at the channel-dielectric interface. Finally, it has been
shown that an underlap can be used to recover the degraded
subthreshold slope and on-off ratio, meeting the intrinsic gate
delay projection by ITRS 2009.

\end{document}